\documentstyle[12pt]{article} \hoffset -0.5in \textwidth 6.5in \textheight
9.00in  \parskip 7pt \openup4.0\jot \parindent=0.5in
in

\def\chain#1#2{\mathrel{\mathop{\null\longrightarrow}\limits^{#1}_{#2}}}

\def\Re{{\cal R \mskip-4mu \lower.1ex \hbox{\it e}\,}}
\def\Im{{\cal I \mskip-5mu \lower.1ex \hbox{\it m}\,}}

\def\etal{{\it et al.}}

\def\tev{\,{\rm TeV}}
\def\gev{\,{\rm GeV}}

\def\mh{\ifmmode m\sbl H \else $m\sbl H$\fi}
\def\mch{\ifmmode m_{H^\pm} \else $m_{H^\pm}$\fi}
\def\mt{\ifmmode m_t\else $m_t$\fi}
\def\mc{\ifmmode m_c\else $m_c$\fi}
\def\mz{\ifmmode M_Z\else $M_Z$\fi}
\def\mw{\ifmmode M_W\else $M_W$\fi}
\def\mr{\ifmmode M_R\else $M_R$\fi}
\def\mws{\ifmmode M_W^2 \else $M_W^2$\fi}
\def\mhs{\ifmmode m_H^2 \else $m_H^2$\fi}
\def\mzs{\ifmmode M_Z^2 \else $M_Z^2$\fi}
\def\mts{\ifmmode m_t^2 \else $m_t^2$\fi}
\def\mcs{\ifmmode m_c^2 \else $m_c^2$\fi}
\def\mchs{\ifmmode m_{H^\pm}^2 \else $m_{H^\pm}^2$\fi}
\def\ztwo{\ifmmode Z_2\else $Z_2$\fi}
\def\zone{\ifmmode Z_1\else $Z_1$\fi}
\def\mtwo{\ifmmode M_2\else $M_2$\fi}
\def\mone{\ifmmode M_1\else $M_1$\fi}
\def\tb{\ifmmode \tan\beta \else $\tan\beta$\fi}
\def\xw{\ifmmode x\sub w\else $x\sub w$\fi}
\def\ch{\ifmmode H^\pm \else $H^\pm$\fi}
\def\lum{\ifmmode {\cal L}\else ${\cal L}$\fi}
\def\inpb{\ifmmode {\rm pb}^{-1}\else ${\rm pb}^{-1}$\fi}
\def\infb{\ifmmode {\rm fb}^{-1}\else ${\rm fb}^{-1}$\fi}
\def\epem{\ifmmode e^+e^-\else $e^+e^-$\fi}
\def\ppb{\ifmmode \bar pp\else $\bar pp$\fi}

\newskip\zatskip \zatskip=0pt plus0pt minus0pt
\def\matth{\mathsurround=0pt}

\def\atversim#1#2{\lower0.7ex\vbox{\baselineskip\zatskip\lineskip\zatskip
  \lineskiplimit 0pt\ialign{$\matth#1\hfil##\hfil$\crcr#2\crcr\sim\crcr}}}

\renewcommand{\thefootnote}{\fnsymbol{footnote}}

\begin{document} \begin{titlepage}
\rightline{\vbox{\halign{&#\hfil\cr
&{\bf OITS-502}\cr
&November 1992\cr}}}
\vspace{1in}
\begin{center}

{\Large\bf
SO(10) Grand Unification with a Low-Energy $M_R$}
\medskip

\normalsize N.G\ Deshpande, E.\ Keith, and T.G.\ Rizzo
\\ \smallskip

Institute of Theoretical Science\\
University of Oregon\\
Eugene, OR\  97403\\
\smallskip

\end{center}

\begin{abstract}

Contrary to commonly held belief, we
show that  one can obtain a low value for $M_R$, the $SU(2)_R$ breaking scale,
in grand unification  theories based on $SO(10)$. This possibility emerges
in the supersymmetric  version of $SO(10)$ with a judicious choice of Higgs
content. The unification scale is found to be consistent with the constraint
from proton decay. This result is first explicitly demonstrated using the
one loop
renormalization group equations, and then a full two loop analysis is
carried out.

\end{abstract}

\renewcommand{\thefootnote}{\arabic{footnote}} \end{titlepage}


It is commonly believed\cite{one,two}  that the group $SU(2)_R$ has to be
broken at a large energy scale $M_R \sim 10^{10}\gev$  if it is to emerge
from a grand unified symmetry such as $SO(10)$.  This is also assumed to be
true for the supersymmetric version of $SO(10)$\cite{one,two}.
Consequently, additional gauge bosons that could possibly be produced at
supercollider energies are thought to originate only from additional $U(1)$
factors which lead to $Z'$ bosons\cite{three}.  The phenomenology of new
charged $W'$ bosons at supercolliders is therefore less frequently
investigated\cite{four}. We will show in this Letter that although the
above result is  true for the simplest Higgs structure,  if this sector
is  suitably enlarged, the scale for the right handed gauge bosons,
$M_R$,  could be made arbitrarily low. We will consider only the
supersymmetric version in detail, with some brief remarks on the
non-supersymmetric case given at the end of our discussion.

We investigate the breaking chain
\begin{equation}
SO(10)\, (SUSY)
\chain{M_U}{}  2_L 2_R 1_{B-L} 3_C\, (SUSY)  \chain{M_R}{}   2_L 1_Y 3_C\,
(SUSY) \chain{M_Z} {}   3_C1_Q\, .\label{+}
\end{equation}
where, as an example of our notation, $2_L$ represents $SU(2)_L$.
Here, we have set the `effective' supersymmetry breaking scale to be $M_Z$, and
will comment on this later.

In Ref.~5, it was shown that if both $2_L 2_R 1_{B-L} 3_C$
singlets of the {\bf 210} representation, together with the $2_L 2_R
1_{B-L}3_C$
singlet of the {\bf 45} representation, acquire vacuum expectation values
(VEVs) then this is sufficient to break supersymmetric $SO(10)$ down to
supersymmetric
$2_L 2_R 1_{B-L} 3_C$ without D-parity.  The breaking at $M_R$ can be
performed
either by the Higgs fields in the ${\bf 126} \oplus \overline{\bf 126}$
representation, or in the ${\bf 16} \oplus \overline{\bf 16}$
representation and we consider both these possibilities in our discussion
below. We further assume that ordinary electroweak breaking at the $Z$ scale is
achieved as usual by a complex {\bf 10} representation. For the purpose of
generating fermion masses, we assume that the entire bi-doublet of the
{\bf 10} representation has a mass at the scale of $M_Z$.  (We remind the
reader that a bi-doublet corresponds to the $(2,2,0,1)$ representation of
$2_L2_R1_{B-L}3_C$.) Also, we assume
that the $SU(2)_R$ triplets of the {\bf 126} and $\overline{\bf 126}$
representations and the $SU(2)_R$ doublets of {\bf 16} and $\overline{\bf 16}$
representations have masses at the scale $M_R$. All other Higgs
multiplets are given masses of order $M_U$ as follows from the survival
hypothesis. We make the important
observation that in this symmetry breaking pattern pseudo-Goldstone
bosons do not appear\cite{five}.

First let us examine the one loop equations:
\begin{eqnarray}
\alpha^{-1}_{1Y}(M_Z) &=&  \alpha^{-1}_{U}(M_U)+{b_{1Y}\over 2\pi}R
+{1\over 2\pi}\left( {3b_{2R}\over 5} +{2b_{B-L}\over 5}\right) (U-R)\, ,
\nonumber\\
\alpha^{-1}_{2L}(M_Z) &=&
\alpha^{-1}_{U}(M_U)+{b_{2L}\over 2\pi}U\, ,\label{++} \\
\alpha^{-1}_{3C}(M_Z) &=&  \alpha^{-1}_{U}(M_U)+{b_{3C}\over 2\pi}U\,
,\nonumber
\end{eqnarray}
where
\begin{eqnarray}
R=\ln{M_R\over M_Z}\, ,\nonumber\\
U=\ln{M_U \over M_Z}\, .
\end{eqnarray}
The $b_i$'s are the one loop beta functions, which for the supersymmetric case
are given by
\begin{equation}
b_{N}^{SUSY}=2 n_g-3N+T(S_N)\, ,
\end{equation}
for $n_g$ generations, the gauge group $SU(N)$, and the complex Higgs fields
contribution which is parameterized by $T(S_N)$. For $U(1)$ gauge groups,
$N=0$ in the above equation and the gauge
couplings are normalized as usual.  Explicitly we find the
Higgs contributions to be given by
\begin{eqnarray}
T_{1Y} &=& {3\over5}n_{10}\; , \; T_{2L}=n_{10}\; , \; T_{3c}=0\, ,\nonumber\\
T_{2R} &=& n_{10}+n_{16}+4n_{126}\; , \; T_{1X}={3\over 2}n_{16}+ 9n_{126}\,
,\label{+++}
\end{eqnarray}
where the subscripts on the $T$'s refer to the relevant
gauge group.  In the above, $n_{10}$ is the number of complex {\bf 10} Higgs
bi-doublets at the scale $M_Z$, and $n_{16}$ and $n_{126}$ are the number
of ${\bf 16} \oplus \overline{\bf 16}$ and ${\bf 126} \oplus \overline{\bf
126}$
Higgs pairs, respectively, which are  used to break the intermediate gauge
symmetry. Using
Eqs\@. (\ref{++}) and (\ref{+++}) together with the definitions
\begin{eqnarray}
\alpha_{1Y}^{-1}(M_Z) &=&  {3\over 5}{1-\tilde{x} \over \alpha(M_Z)}\,,
\nonumber \\
\alpha_{2L}^{-1}(M_Z) &=& {\tilde{x} \over \alpha (M_Z)}\, ,
\end{eqnarray}
gives the relations
\begin{eqnarray}
{2\pi\over \alpha (M_Z)}\left(1-{8\alpha (M_Z)\over 3 \alpha_{3C}(M_Z)}
\right) &=&  \left( C_1-C_2\right) U+C_2 R\, ,\nonumber\\
{2\pi\over \alpha(M_Z)}\left( 1-{8\over 3}\tilde{x} \right) &=& \left(
{5\over 3}C_3-C_2\right) U+C_2 R\, , \label{++++}
\end{eqnarray}
where the abbreviation $\tilde{x} \equiv \sin^2 \theta_W (\overline{MS})$
is used, and the $C_i$ are given by
\begin{eqnarray}
C_1 &\equiv &b_{2L}+{5\over 3}b_{1Y}-{8\over 3}b_{3C}=18+2n_{10}\, ,\nonumber\\
C_2 &\equiv &{5\over 3}b_{1Y}-b_{2R}-{2\over 3}b_{1X}=6-2n_{16} -10n_{126}\,
,\label{XX}\\
C_3 &\equiv & b_{1Y}-b_{2L}=6-{2\over 5}n_{10}\, .\nonumber
\end{eqnarray}
We make the observation that if $C_2=0$ then the scale
$M_R$ is completely undetermined at the one loop level. This gives us
hope that when $C_2=0$, a solution with a low energy $M_R$ will exist.
We can have $C_2=0$ only when $n_{16}=3$ and $n_{126}=0$. We then need only
to require that the two equations in (7) give agreeing values of
$M_U$ within the level of accuracy of the one loop approximation. The
latest values\cite{six} of the input parameters that we use in this analysis
are
\begin{eqnarray}
\alpha^{-1}(M_Z) &=& 127.9\pm 0.1\, ,\nonumber\\
\alpha_{3C} (M_Z) &=& 0.118\pm 0.007\, ,\label{X} \\
\tilde{x} (M_Z) &=& 0.2326\pm 0.0011\, ,\nonumber\\
M_Z &=& 91.187\pm 0.007\, GeV\, .\nonumber
\end{eqnarray}
Consistency of the two equations in (7) with  $n_{16}=3$,
$n_{126}=0$ and $n_{10} = 1$, and taking the central values of
$\alpha^{-1} (M_Z)$ and $\tilde{x} (M_Z)$ from above implies
$\alpha_{3C} (M_Z) = 0.112$, which is within the experimentally allowed range
given above.  Of course we will have to perform a full two loop analysis to
insure a solution exists with a low energy $M_R$ with $n_{16}=3$,
$n_{126}=0$ and $n_{10}=1$. Note that for this particular chioce of Higgs
representations the combination
${3\over 5}\alpha^{-1}_{2R} +{2\over 5}\alpha^{-1}_{1X}$ runs identically
at the one loop level as $\alpha^{-1}_{1Y}$ in the minimal
supersymmetric standard model (MSSM). Since supersymmetric $SU(5)$
grand unification is consistent with $M_S \simeq M_Z$\cite{one,two,seven}, one
would naively expect that a two loop analysis will show the $SO(10)$ scenario
to be equally consistent and we will now show this explitcitly.

We numerically integrate the two loop equations
\begin{equation}
\mu{\partial \alpha_i (\mu)\over \partial \mu}={1\over 2\pi} \left(
b_i+{b_{ij}\over 4\pi}\alpha_j (\mu) \right) \alpha_i^2 (\mu) \, .\label{+++++}
\end{equation}
assuming the breaking pattern as given in Eq.~(1) (with $n_{10}=1,\
n_{16}=3$ and $n_{126}=0$) and we use the
approximation that all superparticles have mass $M_S=M_Z$. Actually,
the `average' sparticle masses could be somewhat different than this effective
value\cite{one,two,seven}. We will also assume  that the top quark and right
handed neutrino
have masses  $\simeq M_Z$. We use the appropriate two loop matching
conditions\cite{eight} at $M_U$ that follows from dimensional reduction:
\begin{equation}
\alpha^{-1}_U
(M_U)-{C_U \over 12\pi}=\alpha^{-1}_i (M_U)-{C_i\over 12\pi} \, ,
\end{equation}
where $i$ represents the intermediate gauge groups $2_L$,
$2_R$,  $1_{B-L}$ or $3_C$ and $C_G$ is the quadratic Casmir invariant
for group $G$. Similarly at $M_R$ we have:
\begin{equation}
\alpha^{-1}_{1Y}(M_R)={3\over 5}\left( \alpha^{-1}_{2R} (M_R)
-{C_2 \over 12\pi}\right) +{2\over 5}\alpha^{-1}_{1B-L} (M_R)\, .
\end{equation}
$C_G=N$ for $SU(N)$ and $C_G=0$ for $U(1)$.
In Eq\@.(\ref{+++++}), we assume the MSSM below the scale \mr\ so that,
 with $i=1_Y \, ,\, 2_L \, ,\, 3_C$, respectively, we use the following two
loop beta functions\cite{nine}
\begin{eqnarray}
b^{MSSM}_{ij}=\left( \matrix{ 0&0&0\cr 0&-24&0\cr
0&0&-54\cr} \right)   +n_g \left( \matrix{  {38\over 15}&{6\over
5}&{88\over 15}\cr {2\over 5}&{14}&{8}\cr {11\over 15}&{3}&{68\over
3}\cr} \right) +n_{10} \left( \matrix{  {9\over 25}&{9\over 5}&{0}\cr
{3\over 5}&{7}&{0}\cr {0}&{0}&{0}\cr} \right) \, .
\end{eqnarray}
For the intermediate symmetry SUSY $2_L 2_R 1_{B-L} 3_C$, we derive from the
generic two loop expression in reference\cite{nine}
\begin{eqnarray}
b^{int.}_{ij} &=& \left( \matrix{ {-24}&{0}&{0}&{0}\cr
{0}&{-24}&{0}&{0}\cr {0}&{0}&{0}&{0}\cr {0}&{0}&{0}&{-54}\cr } \right)
+n_g \left( \matrix{ {14}&{0}&{1}&{8}\cr {0}&{14}&{1}&{8}\cr
{3}&{3}&{7\over 3}&{8\over 3}\cr {3}&{3}&{1\over 3}&{68\over 3}\cr }
\right)  \nonumber\\ &+& n_{10} \left( \matrix{ {7}&{3}&{0}&{0}\cr
{3}&{7}&{0}&{0}\cr {0}&{0}&{0}&{0}\cr {0}&{0}&{0}&{0}\cr } \right)
+n_{16} \left( \matrix{ {0}&{0}&{0}&{0}\cr {0}&{7}&{3\over 2}&{0}\cr
{0}&{9\over 2}&{9\over 4}&{0}\cr {0}&{0}&{0}&{0}\cr } \right) \, ,
\end{eqnarray}
where $i\, ,\, j=2_L\, ,\, 2_R\, ,\, 1_{B-L}\, ,\, 3_C$
respectively. In evaluating the above two equations for $b_{ij}$, we of course
assume
$n_{g}=3$, $n_{10}=1$  and $n_{16}=3$. We treat {\mr } as a free parameter.
{}From the one loop calculation, we expect that the
two loop analysis will yield solutions for arbitrary values of $M_R$ between
$M_Z$ and $M_U$. We have explored the possibility that {\mr } can take on a
wide range of potential values and find the expectation above to be fulfilled.
As an example of this and in particular to show that
$M_R$ can be low, we display in Fig. 1 the case $M_R = 1\tev$. We find that
$\alpha^{-1}_{U} (M_U)=23.4\pm 0.5$   and $M_U =10^{16.2 \pm .4}\gev$,
which is sufficiently large to be consistent with the non-observation of
proton decay. We also note that at the scale $M_R$,  $\alpha_{2L} / \alpha_{2R}
\simeq 1.5$, thus implying the bound $M_{W_R} > 380\gev$ from muon
decay\cite{ten}, and 480 GeV from direct collider searches\cite{eleven,twelve}.
We have investigated the influence of heavy
top quark Yukawa couplings on the two loop evolution\cite{seven}, and find the
effect to be negligible. {\it Thus, at least for this choice of symmetry
breaking, {\mr } can indeed be sufficiently low as to be of consequence for
existing and future colliders.}

We have now described the conditions under which $M_R$ may be associated
with a low
scale in supersymmetric $SO(10)$ grand unification. From the previous
discussion, we can see that we need $M_S \simeq M_Z$ for our scenario to be
realized as in the case of SUSY $SU(5)$. What about non-supersymmetric
$SO(10)$ grand unification? In this case, the
equations analagous to (7) and (8) imply
that for $M_R$ to drop out of the one loop equations we require that
$5n_{126}+n_{16}=22$, where $n_{126}$ and $n_{16}$ refer to the number
of {\bf 126} and {\bf 16} dimensional representations, respectively, used to
break the intermediate gauge symmetry $2_L 2_R 1_{B-L} 3_C$ to the SM.
Demanding consistency of
the two one loop equations analagous to Eq.~(8) and using
the  values of the low energy parameters as given above
further implies $n_{10} = 4$, where $n_{10}$ is the number of scalar
bi-doublets at the scale $M_Z$. (The entire Higgs multiplet $(2,2,0,1)$,
within the {\bf 10}
representation must have mass less than $M_R$, otherwise the one loop
equations will still depend on $M_R$.) This illustrates the result that
in order to achieve grand unification in this  non-supersymmetric
$SO(10)$ case, as in the conventional $SU(5)$ model\cite{seven},
we would have to employ many Higgs doublets. In the $SO(10)$ case, we then
obtain unification with $M_U\simeq 10^{13.6}\gev$, a value which is clearly
inconsistent with limits on the proton lifetime.

In summary, we have shown for the first time that a low $SU(2)_R$ breaking
scale is
compatible with $SO(10)$ grand unification. The symmetry breaking at the
scale $M_R$ is accomplished by three generations of Higgs in the {\bf 16}
representation, not unlike the three families of quarks and leptons.
Further consequenses of this symmetry breaking scenario for neutrino masses
will be the subject of a future investigation.

\vskip.25in
\centerline{ACKNOWLEDGEMENTS}

TGR would like to thank the Institute of Theoretical Science for its
hospitality and JoAnne L. Hewett for discussions related to this work. This
work was supported in part by the U.S. Department of Energy grant number
DE-FG06-85ER-40224.

\newpage

%
\def\MPL #1 #2 #3 {Mod.~Phys.~Lett.~{\bf#1},\ #2 (#3)}
\def\NPB #1 #2 #3 {Nucl.~Phys.~{\bf#1},\ #2 (#3)}
\def\PLB #1 #2 #3 {Phys.~Lett.~{\bf#1},\ #2 (#3)}
\def\PR #1 #2 #3 {Phys.~Rep.~{\bf#1},\ #2 (#3)}
\def\PRD #1 #2 #3 {Phys.~Rev.~{\bf#1},\ #2 (#3)}
\def\PRL #1 #2 #3 {Phys.~Rev.~Lett.~{\bf#1},\ #2 (#3)}
\def\RMP #1 #2 #3 {Rev.~Mod.~Phys.~{\bf#1},\ #2 (#3)}
\def\ZP #1 #2 #3 {Z.~Phys.~{\bf#1},\ #2 (#3)}
\def\IJMP #1 #2 #3 {Int.~J.~Mod.~Phys.~{\bf#1},\ #2 (#3)}

\newpage

\noindent{\bf Figure Caption}
\begin{itemize}
\item[Figure 1.]{Evolution of coupling constants for the symmtery breaking
chain given in Eq.~(1). We have used $n_{10}=1$, $n_{16}=3$,
$n_{126}=0$ and $M_R=1\tev$, where these quantities are defined in the
text. The error bars we show arise from  uncertainties in the low
energy parameters in Eq. (9). In the figure,
$\alpha^{-1}_{i}$ is calculated via the dimensional reduction scheme at two
loop order.\label{fig1}}
\end{itemize}

\end{document}